\documentclass[journal]{IEEEtran}

\usepackage{url}  

\usepackage[noadjust]{cite}
\usepackage{enumitem}
\usepackage{amsmath,amssymb,bm}
\usepackage{booktabs}
\usepackage{graphicx,epstopdf,subfigure}
\usepackage{multirow}
\usepackage{tabularx,threeparttable}
\usepackage{array}
\usepackage{xcolor}
\usepackage{comment}
\usepackage{makecell}
\usepackage[normalem]{ulem} 
\usepackage[colorlinks=true,citecolor=blue,urlcolor=blue]{hyperref}
\usepackage{cite}

\usepackage{fancyhdr}

\begin{document}

\title{Efficient Medical Image Segmentation Based on Knowledge Distillation}

\author{Dian Qin, Jia-Jun Bu, Zhe Liu, Xin Shen, Sheng Zhou, Jing-Jun Gu, Zhi-Hua Wang, Lei Wu, Hui-Fen Dai

\thanks{This work is supported by the National Natural Science Foundation of China (Grant No. 61972349), Soft Science Research Project of Zhejiang Province Science and Technology Department (2020C25035) and Key Research and Development Program of Zhejiang Province (No. 2018C03085 and 2021C03121) \emph{(Corresponding author: Jia-Jun Bu)}}
\thanks{Dian Qin, Jia-Jun Bu, Zhe Liu, Xin Shen, Jing-Jun Gu, Zhi-Hua Wang, and Lei Wu are with Zhejiang Provincial Key Laboratory of Service Robot, College of Computer Science, Zhejiang University (e-mail: qindian@zju.edu.cn; bjj@zju.edu.cn; zheliu@zju.edu.cn; xinshen@zju.edu.cn; gjj@zju.edu.cn; zhihua\_wang@zju.edu.cn; shenhai1895@zju.edu.cn).}
\thanks{Sheng Zhou is with Ningbo Research Institute, School of Software Technology, Zhejiang University (e-mail: zhousheng\_zju@zju.edu.cn).}
\thanks{Hui-Fen Dai is with The Fourth Affiliated Hospital Zhejiang University School of Medicine (e-mail: daihuifen@zju.edu.cn).}}

\markboth{IEEE Transactions on Medical Imgaging}
{Shell \MakeLowercase{\textit{et al.}}: Bare Demo of IEEEtran.cls for Journals}
	
\maketitle


\begin{abstract}
Recent advances have been made in applying convolutional neural networks to achieve more precise prediction results for medical image segmentation problems. However, the success of existing methods has highly relied on huge computational complexity and massive storage, which is impractical in the real-world scenario. To deal with this problem, we propose an efficient architecture by distilling knowledge from well-trained medical image segmentation networks to train another lightweight network. This architecture empowers the lightweight network to get a significant improvement on segmentation capability while retaining its runtime efficiency. We further devise a novel distillation module tailored for medical image segmentation to transfer semantic region information from teacher to student network. It forces the student network to mimic the extent of difference of representations calculated from different tissue regions. This module avoids the ambiguous boundary problem encountered when dealing with medical imaging but instead encodes the internal information of each semantic region for transferring. Benefited from our module, the lightweight network could receive an improvement of up to 32.6\% in our experiment while maintaining its portability in the inference phase. The entire structure has been verified on two widely accepted public CT datasets LiTS17 and KiTS19. We demonstrate that a lightweight network distilled by our method has non-negligible value in the scenario which requires relatively high operating speed and low storage usage.
\end{abstract}


\section{Introduction}

\begin{figure}[t]
\centering
\includegraphics[width=0.8\columnwidth]{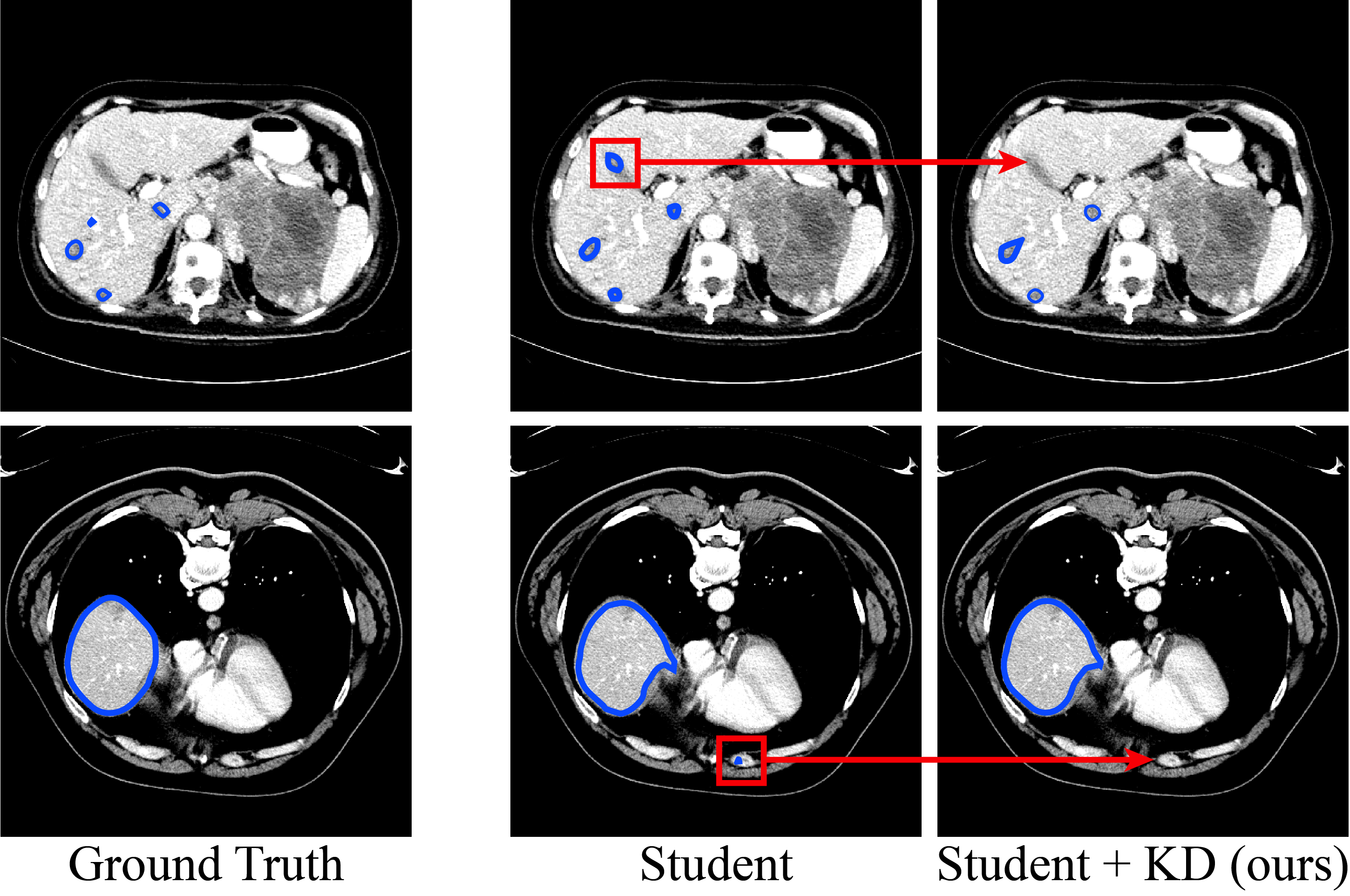}
\caption{Experimental results on LiTS. The first row represents a case of liver tumor segmentation and the second row is from liver segmentation experiments. The red arrows indicate the powerful error correction capability of the method we proposed.}
\label{figure_detail}
\end{figure}

\IEEEPARstart{M}{edical} image segmentation aims to provide pixel-level semantic interpretation by generating segmentation masks of organs and tumors automatically. However, some organic characteristics such as diverse appearances, irregular sizes, unpredictable locations, and different variations with the contrast agent make medical image segmentation more challenging than the semantic segmentation on daily photographic pictures. Deep learning has been introduced to the field to deal with these problems. Some methods such as convolutional neural network (CNN) are first applied in medical image processing in a relatively straightforward way. Two representative examples are CNN with graph cut \cite{cut} and CNN with conditional random fields \cite{random}. The successful practice of quite a few medical image segmentation challenges such as the liver tumor segmentation challenge (LiTS) \cite{lits}, the Kidney Tumor Segmentation Challenge (KiTS) \cite{kits}, and the Multimodal Brain Tumor Image Segmentation Challenge (BraTS) \cite{brats} simulates the break out of the researches for solving biomedical segmentation using convolutional networks. 

With the appearance of UNet \cite{unet}, many efforts have been made in medical image segmentation methods, such as adding dense connections, replacing the feature extractors, and adopting 3D convolution kernels. For examples, the model RA-UNet \cite{rau} incorporates the attention mechanism \cite{attention} based on UNet architecture. H-DenseUNet \cite{hdu} has an eye-catching performance in the LiTS challenge with a hybrid use of the DenseNet \cite{dense}, UNet structure, and volumetric information. Some other researches realize the importance of capturing information of spatial continuity, they directly expand the dimension of convolution kernel from 2D to 3D such as the network 3D U-Net \cite{3dunet} and 3D U$^2$-Net \cite{3du2net}. However, the methods mentioned above are inevitable to append various expensive computation components and enlarge the required storage. It is increasingly difficult to deploy in real-world scenarios. Although, a large number of researches such as ENet \cite{enet} and ERFNet \cite{erf} about lightweight networks have been applied in real-time semantic segmentation. Some recent works \cite{realtime} also have started paying attention to real-time medical image segmentation problems. There is still a dilemma that the performance tends to be damaged when the models are simplified for faster speed.

To overcome the above limitations of existing methods, technics including model compression, transfer learning, and knowledge distillation \cite{kd} are introduced. Among them, knowledge distillation has attracted broad attention from both academics and industries. It tries to distill information from a well-trained teacher network to another lightweight student network to improve the performance of the latter. The original distillation methods can only transfer the logits of the final convolution layer as information. However, the information in the learning process is largely ignored. Recently, some efforts have been made in semantic segmentation field to deal with this problem by disposing intermediate features. For example, the knowledge adaption for segmentation \cite{akd} adopts self-supervised learning to translate knowledge from the teacher network. The structured knowledge distillation \cite{skd} comprises pair-wise distillation technology and generative adversarial learning to distill holistic semantic information. The intra-class feature variation distillation (IFVD) \cite{ifvd} presents the idea of calculating intra-class similarities among pixels with the guidance of labeled segmentation masks. Unfortunately, the above methods have not considered the effectiveness of distillation in medical image scenarios.

Only a few researches have studied the efficiency of segmentation for medical imaging problems and utilized the knowledge distillation technology in recent years. Two pioneer works apply knowledge distillation in dealing with chest X-Ray \cite{xray} and 3D optical microscope images \cite{3dopt}. Most subsequent researches have focused on multi-modal problems. For example, mutual knowledge distillation \cite{mkd} is proposed to deal with the cross-modality problem for different computerized tomography (CT) and magnetic resonance imaging (MRI) scans with the same semantic information. The method devised by \cite{umkd} brings knowledge distillation into unpaired multi-modal segmentation to reach good performance. The work in \cite{mono} tries to distill knowledge from multi-modal to mono-modal segmentation networks. However, the above methods either ignore the intermediate features or require fixed networks for distillation. To the best of our knowledge, almost no method considering to construct a systematical knowledge distillation architecture for the general and single-modal medical image segmentation problems so far. The reason could be that it is challenging to explicitly extract features that are conducive to segmentation from complicated medical images.

In this paper, we discuss dealing with the above problems by introducing a holistic and robust architecture with a novel core module that is custom-made for encoding and transferring region information in medical imaging. First, we propose a distillation architecture that can excavate information from off-the-shelf medical image segmentation networks and transfer them to another lightweight network called the student network. Then, we devise the Region Affinity Distillation (RAD) module to encode and distill the importance of semantic region information in medical imaging segmentation scenarios. Concretely, the collection of inter-class contrasts between different tissue regions, dubbed region contrast map, is calculated from intermediate feature maps with the guidance of ground truth segmentation masks. The RAD module forces the student network to mimic its teacher in terms of the region contrast map to learn the segmentation capability indirectly. This new module avoids the ambiguous boundary problem encountered when dealing with medical imaging but instead encodes the internal information of each semantic region. Figure \ref{figure_detail} shows that the effectiveness of our method is strong enough to correct some subtle segmentation errors produced by the student network.

Extensive experiments conducted on public datasets LiTS and KiTS demonstrate the remarkable performance of our method. The student model distilled by our method can improve up to 32.6\% from the dice coefficient of 0.516 to 0.684 for the tumor segmentation in our experiments. This improvement is remarkable while looking at the entire field of semantic segmentation. Our method can also narrow the performance gap between the teacher network and the student network nearly 3.75 times, that is, from 0.229 to 0.061. Note that the size of this student network is 21 times smaller than his teacher. It makes it possible that the lightweight methods can be the alternatives for cumbersome networks in most real-world scenarios of medical image segmentation in the future.

Overall, we summarize our contributions as follows.

(1) We proposed a knowledge distillation based architecture that systematically constructs a holistic structure for transferring segmentation capability when processing with medical imaging.

(2) We devised a novel Region Affinity Distillation (RAD) module that aims to encode regional knowledge for student networks to mimic, which is essential to improve the segmentation performance when dealing with medical imaging by being aware of the difference of semantic information among regions.

(3) We demonstrated the feasibility and reproducibility through robust experiments on two public medical image datasets LiTS and KiTS19 with sufficient ablation considerations.

\section{Related Work}
\label{sec:related}
\subsection{Medical Image Segmentation}
The last few years have witnessed a sustainable development of researches about the medical image segmentation problem. UNet family \cite{unet}-\cite{3du2net} is known as an effective architecture that can address medical imaging problems \cite{liseg}-\cite{brseg}. Benefited from the straightforward semantic information and relatively stationary imaging structure, the skip-connection of UNet or its familial networks leads the decent performance most of the time. The utilization of variants of the generative adversarial network (GAN) \cite{syngan}-\cite{higan} aroused recently. The Radiomics-guided Gan \cite{rggan} aims to generate segmentation of tumor from non-contrast images by fusing the radiomics feature of contrast CT images as prior knowledge. Training networks with the adversarial strategy \cite{ligan} seems to be another way to adopt the GAN mechanism. Moreover, semantic segmentation methods have always received medical imaging researchers' attention, such as PSPNet \cite{psp} and Deeplab series networks \cite{deep}. The models mentioned above are suitable to be assigned the role of teachers in our architecture, as they have well performance but relatively high requirements for storage and computation. 

Besides, we need some lightweight segmentation networks to play the role of students. Although some researches on the lightweight network for medical image have appeared recently, such as SA-UNet \cite{sau} and lightweight attention CNN \cite{lacnn} for retinal vessel segmentation, there is no widely accepted lightweight model dedicated to medical image segmentation so far. In practice, the full-convolution-based lightweight networks are more commonly adopted in various segmentation scenario, such as ENet \cite{enet}, ESPNet \cite{esp}, ERFNet \cite{erf}, ShuffleNet \cite{shuffle}, SqueezeNet \cite{sqnet}, and MobileNet \cite{mv2}. In this paper, we implement some of these methods and make them the students in our distillation architecture.

\subsection{Knowledge Distillation}
Knowledge Distillation \cite{kd} is an approach of transferring knowledge from a powerful but cumbersome network to the lightweight model to improve the performance of the latter without affecting its efficiency. Many researchers \cite{kd}-\cite{crd} utilized it to deal with classification problems by distilling knowledge from the output class probabilities of excellent models. Feature normalized knowledge distillation \cite{fnkd} gives a good example of optimizing the metric function between the logits exported from the teacher and student networks.

The method proposed by \cite{at} further guides the compact networks to mimic intermediate features extracted from pre-trained teacher network by constructing attention maps. Similarity-preserving knowledge distillation \cite{spkd} proposes a fresh distillation structure by measuring the similarity between samples. Recently a batch of knowledge distillation methods aroused for handling the object detection and semantic segmentation problems \cite{akd}-\cite{mdk}. They are devoted to exploring available approaches to distill interior structural information that can benefit the segmentation task in theory. Exceptionally, mutual knowledge distillation \cite{mkd} was proposed to solve the multimodal medical imaging problems by learning segmentation abilities from each other. We conduct the novel distillation architecture in this paper based on some of the methods mentioned above.

\section{Methodology}
\label{sec:method}

\begin{figure*}[t]
\centering
\includegraphics[width=1\textwidth]{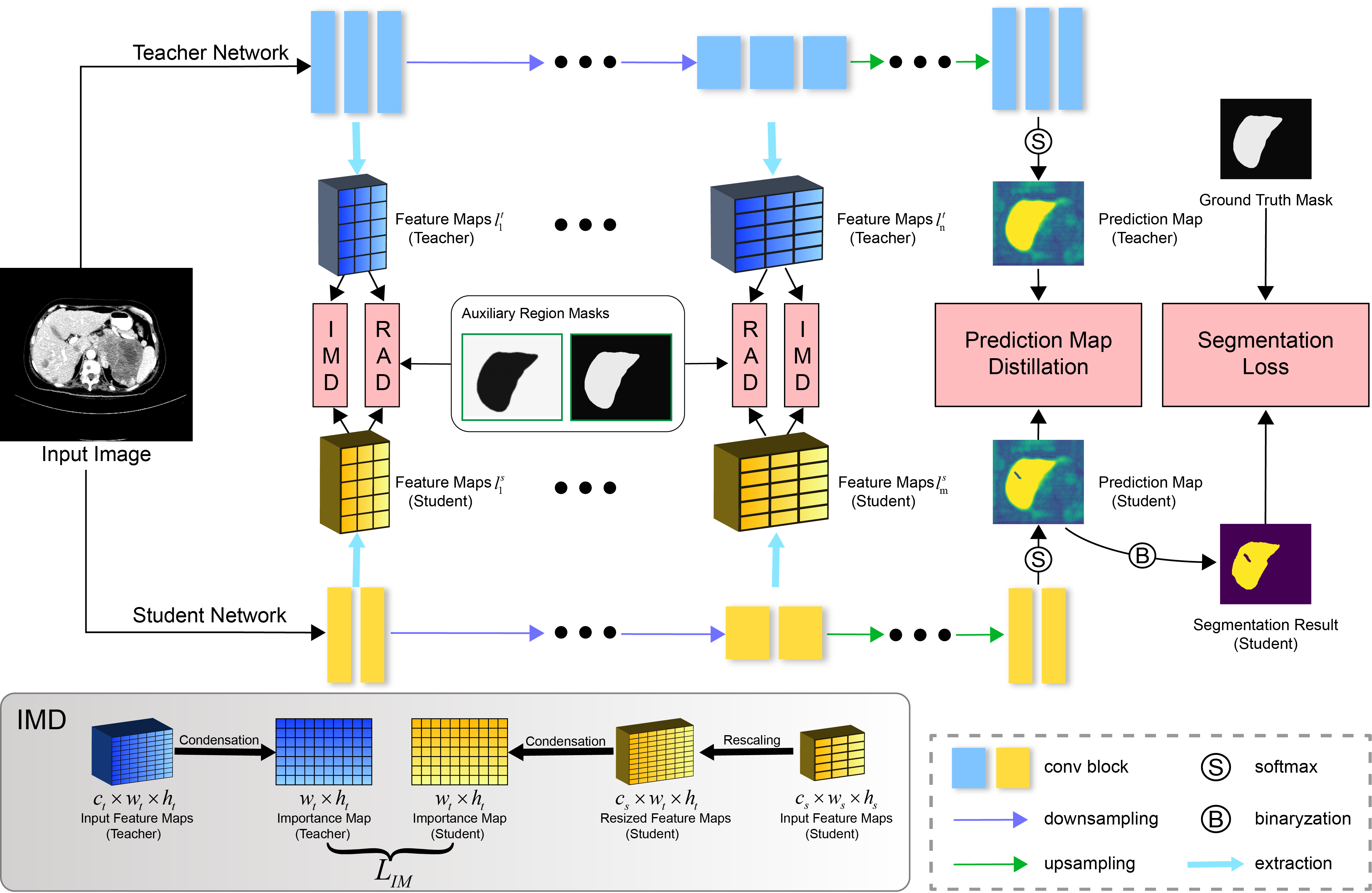}
\caption{The pipeline of proposed distillation architecture. The teacher network and the student network are represented by two horizontal path lines up and down. They take the same images as input simultaneously and output their own predictions. As the area shown between the two networks, we divide the distillation process into several blocks which are in charge of the distillation process and the segmentation task. The Importance Maps Distillation (IMD) module, Region Affinity Distillation (RAD) module, and the Prediction Map Distillation (PMD) module carry the knowledge distillation mechanism in our structure. In particular, the RAD module needs extra inputs, that is, the auxiliary region masks placed in the middle of the picture. The inner structure of the IMD is illustrated in the left bottom, and the RAD module is in Figure \ref{figure_ra}.}
\label{figure_arc}
\end{figure*}

\begin{figure*}[t]
\centering
\includegraphics[width=0.95\textwidth]{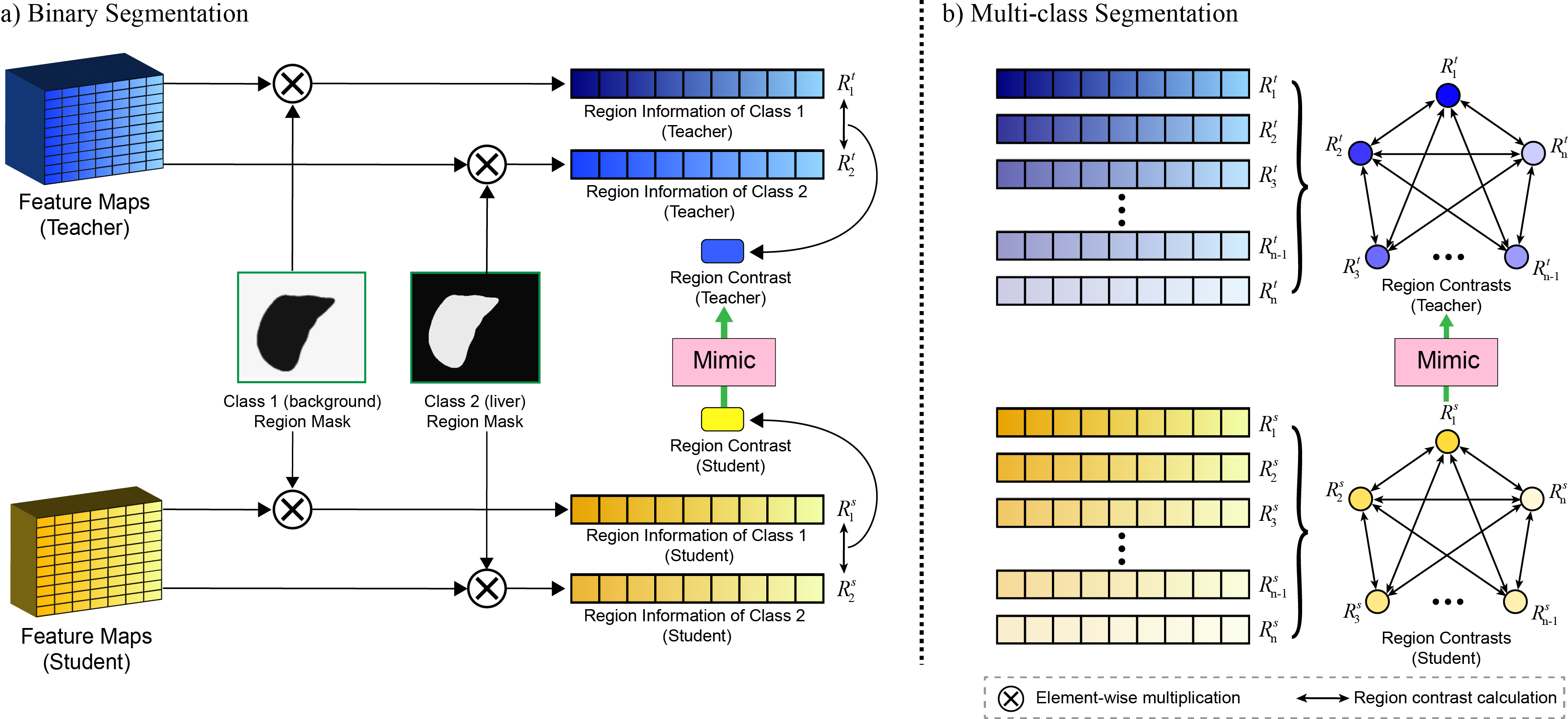}
\caption{The architecture of Region Affinity Distillation (RAD) module. This module accepts teacher feature maps and student feature maps as the input simultaneously and calculates the values of region contrast separately by multiplying the label masks that corresponding to the input. The region affinity loss can be computed with the contrasts in the end. The resized label masks need to be processed by the one-hot operation before the multiplication. We divide the segmentation scenarios into two cases: a) Binary segmentation, as shown in the left part, is the most commonly adopted flow in medical image segmentation problems where the only one type of object needed to be recognized; b) Multi-class segmentation, as illustrated in the right, is another scenario when handling with over 2 types of semantic targets.}
\label{figure_ra}
\end{figure*}

In this section, we decompose the proposed method in detail. The pipeline of the distillation architecture devised by us is illustrated in Figure \ref{figure_arc}. It takes a grayscale CT image of size $W \times H$ as input and exports a segmentation result of the same size. The holistic distillation structure comprises four core modules marked as pink rectangle in the figure. From left to right, the first two modules IMD and RAD take charge of transferring intermediate information by constructing the form of importance maps and region affinity maps respectively. Then, the Prediction Map Distillation module aims to drive the student network to mimic the output of the final layer of the teacher to learn segmentation capability quickly. In the last, it is necessary to append the segmentation task loss to ensure a basic performance corresponding with the domain of inputs. Benefited from this architecture, the student network can take care of its own segmentation task as well as distill experience from the teacher simultaneously. The details of each module are described below.

\subsection{Prediction Maps Distillation}
The basic methodology of knowledge distillation \cite{kd} attempts to drive the student network to acquire knowledge from the teacher network by calculating the difference of their final layer, i.e. the output logits with some measurement functions such as cross entropy and Kullback-Leibler divergence.

Inspired by the distillation method mentioned above, we follow part of the prior works about knowledge distillation for semantic segmentation \cite{akd} \cite{skd} to construct the Prediction Map Distillation module. This module is introduced to enable the student network to learn predictive capability from the output segmentation map of the teacher network explicitly. Here we view the segmentation map as a collection of pixel-level classification problems. Specifically, we calculate a loss value for all pixel pairs at the same spatial position in the two networks, then assemble these values as the distillation loss of this module. The loss function is given as:
\begin{equation}
    L_{PM} = {\frac 1 N} {\sum_{i \in N} {\rm {KL}}(p_i^s || p_i^t)}
\end{equation}
where $N = W \times H$ is the number of pixels of the segmentation map, ${\rm {KL}(\cdot)}$ is the Kullback-Leibler divergence function. $p_i^s$ and $p_i^t$ represent the probabilities of the $i$th pixel in the segmentation map extracted from the student and the teacher network respectively. This module is illustrated as the 2nd pink rectangle from right to left in Figure \ref{figure_arc}.

\subsection{Importance Maps Distillation}
In addition to distilling knowledge from answers, learning the problem-solving process is also an important ability for student networks. For neural networks, the main obstacle is that the sizes of features among the teacher and the student network are usually completely different. To solve this, we introduce the Importance Maps Distillation (IMD) module to encode the feature maps among neural networks into a transformable form. 

The detailed structure of this module is depicted in the bottom left corner of Figure \ref{figure_arc}. Specifically, given the feature maps $e_s$ of size $c_s \times w_s \times h_s$ extracted from an arbitrary layer of the student network and the feature maps $e_t$ of size $c_t \times w_t \times h_t$ extracted from the relatively same location of the teacher network, we first apply a step of rescaling to force the student's feature maps $e_s$ to match the teacher's $e_t$ in spatial scale. This step can be defined as: 
\begin{equation}
    \hat{e}_s = f(e_s); \hat{e}_s \in R^{c_s \times w_t \times h_t}
\end{equation}
The adoption of the rescaling method $f(\cdot)$ depends on the spatial size relationship of $e_s$ against $e_t$, i.e. $w_s \times h_s$ against $w_t \times h_t$, to employ unpooling when smaller, pooling when bigger, and no operation when same.

Then we follow the works of attention transfer \cite{at} with the assumption that the absolute value of a neuron activation indicates the importance of itself. In detail, considering the feature maps $\varepsilon$ of size $C \times w \times h$, we simply sum the absolute value of $\varepsilon$ along the channel dimension $C$ to generate the importance map $M \in R^{w \times h}$ w.r.t. the original features $\varepsilon$. The process is defined as:
\begin{equation}
    \varphi(\varepsilon) = {\sum_{i=1}^C {|\varepsilon_i|}^2}
\end{equation}
where $\varepsilon_i$ denotes the $i$th matrix of $\varepsilon$ along the channel dimension $C$.

Thus, it is possible to distill knowledge by exporting their importance maps. The distillation loss of this module can be calculated by:
\begin{gather}
    M_i^s = \varphi(f(e_i^s)), M_j^t = \varphi(e_i^t) \\ 
    \label{eq_im}
    L_{IM} = {\sum_{(i, j) \in P} ||{\frac {M_i^s} {||M_i^s||_2}} - {\frac {M_j^t} {||M_j^t||_2}}||_1}
\end{gather}
where $e_i^s$ and $e_j^t$ represent the feature maps of $i$th and $j$th layer extracted from the student and the teacher network respectively, $M_i^s$ and $M_j^t$ are their importance maps. $P$ is the collection of the indices pairs of all possible position with the same size of embeddings, and $(i, j)$ is a sample from $P$. Operations $||\cdot||_1$ and $||\cdot||_2$ are the $l_1$ and $l_2$ normalization. Note that the $l_1$ norm is introduced as the importance maps are relatively sparser in medical image segmentation scenarios.

Obviously, compared with the work in \cite{at} that requires the strictly identical spatial size of feature maps of teacher and student networks, our module makes the distillation feasible between feature maps of totally different sizes through an extra simple but practical rescaling process.

\subsection{Region Affinity Distillation}
It is common sense that segmentation models will perform better while realizing the implicit structural information that is easier to capture by cumbersome networks benefited from the deep convolutional layers and large receptive fields. Therefore, when considering constructing a distillation method, the most crucial issue is how to transfer the implicit structural information to lightweight networks. Although the indistinct boundary problems in many tumor segmentation tasks make the distillation very challenging, we still noticed that the difference in the graphic appearance between different semantic regions in CT images is pronounced. Follow this idea, we propose a novel distillation module named Region Affinity Distillation (RAD) by transferring the relationship information between regions from the teacher network to the student network.

To this end, we utilize the labeled segmentation masks which comprise precise areas of every semantic class to extract the region information by classes from feature maps. Then we calculate the region contrast value by measuring the similarity among the region information of these classes. Figure \ref{figure_ra} shows the architecture of RAD module. In detail, let a stack of feature maps extracted from a certain intermediate layer be $\varepsilon$ with the size of $C \times w \times h$. First, we resize the binary label masks $m$ from $W \times H$ to $w \times h$ as the size of feature maps $\varepsilon$ are different from the input image in common. Then, given a semantic class $i$, we can calculate the region information vector $R_i$ of class $i$ by averaging all the features of length $C$ in $\varepsilon$ where the pixel is located in the area that covered by the $i$th binary label mask $m_i$. This process can be implemented by element-wise multiplication as:
\begin{equation}
    R_i = {\frac 1 {N_i}} {\sum_{j=1}^{w \times h}  \varepsilon_j \cdot m_{ij}}
\end{equation}
where $i = 1, 2, ..., c$ is the index of classes, $j$ is the index of pixels of resized shape,  $N_i$ is the number of pixels of valid areas in $i$th mask $m$.  Then, the region contrast value can be computed as:
\begin{equation}
    V_{rc} = {\frac 1 n}{\sum_{(i, j)} {\frac {R_i^{\rm T}R_j} {||R_i||_2||R_j||_2}}}
    \label{eq_sim}
\end{equation}
where $(i, j)$ is a pair of indices among classes, $n$ is the number of all possible class pairs. Note that $V_{rc}$ can also be a vector that comprises all the similarity values before the averaging. In practice, most existing medical image segmentation tasks require only a few objective classes. So that, Eq. \ref{eq_sim} gives a more concise and efficient calculation as it contributes similar effects with the vector form of $V_{rc}$ in general. However, it is reasonable to adopt vector form when facing segmentation problems with numerous semantic classes.

Finally, given the region contrast value/vector $V_{rc}^s$ and $V_{rc}^t$ for the student and the teacher network respectively, the region affinity loss can be calculated by the loss function defined as:
\begin{equation}
    L_{RA} = {\sum_{(i, j) \in P}||V_{rc}^s - V_{rc}^t||_p }
\end{equation}
where $p$ is the norm type, which can be assigned to 1 or 2. The meaning of $i$, $j$, and $P$ is similar to Eq. \ref{eq_im}.

Figure \ref{figure_ra}-a presents the commonly faced binary segmentation scenario ($c=2$) in medical image segmentation. The student network is only asked to mimic the region contrast between the region information of our target area and the background area. When facing with multi-class segmentation problems ($c>2$), one can refer to the Figure \ref{figure_ra}-b. In this case, the student network must mimic the region contrasts graph of the teacher like the polygon illustrated in the rightmost of the figure, which is calculated between the region information of all the possible class pairs.

\subsection{Training Process}
As illustrated in Figure \ref{figure_arc}, we integrate the distillation modules mentioned above to train the student network in an end-to-end manner. The total loss function is given as:
\begin{equation}
    \label{eq_all}
    L_{total} = L_{seg}+ \alpha L_{PM} + \beta_1 L_{IM} + \beta_2 L_{RA}
\end{equation}
where $L_{seg}$ is the general segmentation loss function that can be either of the cross entropy loss and the dice loss \cite{vnet}. The hyper-parameters $\alpha$ is set to 0.1, $\beta_1$ and $\beta_2$ are both set to 0.9. In practice, we always set $\beta_1$ and $\beta_2$ to the same value as our experiments have demonstrated the insensitivity of the value fluctuation of any single one. Check the corresponding experimental results in Sec. \ref{sec:experiments}-D for more details.

Given a well pre-trained teacher network, we train this end-to-end architecture and update the parameters of the student network according to the loss function Eq. \ref{eq_all}. Notice that extracting two to four pairs of representative low-level features and high-level features when use IMD and RAD module to distill process information is the most efficient choice, while all the pairs of features with the same size are available in practice.

The teacher network part and distillation modules will be discarded in the inference phase after training sufficiently. What has been proven by our experiments is that our method can not only gift the lightweight network remarkable improvements but also maintain the number of its parameters.

\section{Experiments}
\label{sec:experiments}    

\subsection{Setup}
To conduct a series of convictive experiments, we adopt state-of-the-art segmentation architectures such as RA-UNet \cite{rau} as the teacher networks and several open-source lightweight networks such as ENet \cite{enet} as the student networks to verify the effectiveness of our distillation method. We follow the official setup including network structures and hyper-parameters when training these architectures solely. All the segmentation networks and distillation processes in our experiments are trained by Adam with the beta1 (0.9) and the beta2 (0.999). The learning rate is initialized as 0.001, and CosineAnnealing is adopted to schedule the learning rate with the lowest learning rate 0.000001. We also employ data augmentation methods such as random rotation and flipping. It has been proven by our experiments that the data augmentation trick of Gaussian noise is not suitable for medical images.

Most networks take the authentic $512 \times 512$ CT images as the input. The HU values of CT images used for input need to be windowed in advance. From the radiology experience, the window width of the CT image is generally set to -40 to 160 for liver, and -200 to 300 for kidney. For the unification of the environment of our experiments, every model used in our experiments was implemented with the Pytorch framework. Algorithms were trained and tested on an NVIDIA GeForce RTX 3090 GPU (24GB). We train all the networks to convergence with up to 60 epochs of training.  We follow the 5-fold cross-validation training strategy and collect the test scores from the last 20 epochs of every fold. Given these collected test scores, all the performance values in our experiments are presented as a format of range value with the form $a \pm b$ where $a + b$ is the maximum and $a - b$ is the minimum.

\subsection{Dataset}
\subsubsection{LiTS}
The most valued LiTS \cite{lits} dataset contains 201 CT scans acquired with different CT scanners and acquisition protocols. As the labeled liver collection of the largest amount of data, scans from LiTS incorporates diverse types of liver tumor disease. The mix of pre-therapy and post-therapy CT images gives the participants a big challenge. The image presentation is very diverse. The image resolution ranges from 0.56mm to 1.0mm in axial and 0.45mm to 6.0mm in z direction. The number of slices in z ranges from 42 to 1026. The size of the tumors varies between 38$\rm {mm^3}$ and 349$\rm{mm^3}$. As the organizers guaranteed the professional level of labeling of both liver and liver tumor, we follow the official split of LiTS, using 131 cases for training and 70 cases for testing. Five-fold cross-validation is adopted in the training process.

\subsubsection{KiTS19}
The publicly accessible KiTS19 \cite{kits} dataset embraces 210 intact abdominal CT scans labeled with manual segmentation masks of kidney and kidney tumor. There is no pre-operative arterial phase data, and the slice thicknesses range from 1mm to 5mm. The image resolution ranges from 0.4mm to 1.0mm in axial. The longitudinal fields of view range from 20 to 140. The volume of most tumors varies between 9.6$\rm{cm^3}$ to 109.7$\rm{cm^3}$. Organizers emphasized that every patient selected into this dataset carries one or more kidney tumors. We simply random sample 168 cases for training and the rest 42 cases for testing. All the pre-processing methods and the operations related to training the networks are the same as the ways used in LiTS.

\subsection{Evaluation Metric}
In medical imaging segmentation problems, the dice coefficient is commonly taken for evaluation. For both applicability and practicality of the volume segmentation task, the mentioned dice score of our experiment means dice coefficient per case uniformly. The metric function of the dice coefficient of a single case is defined as:
\begin{equation}
{\rm {DICE}}(P, G)={\frac {2|P \cap G|} {|P| + |G|}}
\end{equation}
where $P$ and $G$ represent the prediction and ground truth of the volumetric tumor mask respectively.  

We also provide two other segmentation metrics the volume overlap error (VOE) and the relative volume difference (RVD) as a reference, while the dice coefficient is still the chief referee. They are given as follows:
\begin{gather}
    {\rm {VOE}}(P, G) = 1 - {\frac {|P \cap G|} {|P| + |G|}} \\
    {\rm {RVD}}(P, G) = {\frac {|P| - |G|} {|G|}}
\end{gather}

It should be emphasized that VOE and RVD are different from the dice coefficient which the larger the value is, the better the network performance is. They are the metric of errors, that is, we hope those values (or the absolute values) are as small as possible. 

\subsection{Ablation Study}

\begin{table*}[t]
\centering
\setlength{\tabcolsep}{10pt}
\renewcommand\arraystretch{1.4}
\caption{Results of our cross experiments between different teacher and student networks on LiTS and KiTS19. The displayed highlights are the highest dice coefficient scores of their column. The unit of the number of parameters is millions marked as M in the chart. Note that n/a is placed when the performance of the teacher is inferior to the student network, knowledge distillation is not applicable in this case theoretically.}
\begin{tabular}{cccccc}
\hline
Method & Liver Tumor Dice & Liver Dice  & Kidney Tumor Dice & Kidney Dice  & \#Params (M) \\ \hline
\multicolumn{6}{c}{Teachers}                                                                                                   \\ \hline
T1: RA-UNet   & 0.685 $\pm$ 0.004      & 0.960 $\pm$ 0.001      & 0.745 $\pm$ 0.003       &  0.970 $\pm$ 0.001 & 22.1                               \\ 
T2: PSPNet                 & 0.640 $\pm$ 0.005       & 0.959 $\pm$ 0.001     & 0.659 $\pm$ 0.007       &  0.968 $\pm$ 0.002       & 46.7                                \\ 
T3: UNet++                       & 0.669 $\pm$ 0.003      & 0.949 $\pm$ 0.001 & 0.644 $\pm$ 0.007       &     0.943 $\pm$ 0.002   & 20.6                           \\ \hline
\multicolumn{6}{c}{Students and their performances distilled from different teachers by our approach}                                                              \\ \hline
ENet          & 0.574 $\pm$ 0.005      & 0.952 $\pm$ 0.001     & 0.521 $\pm$ 0.015       &    0.939 $\pm$ 0.001       & \multirow{4}{*}{0.353}                         \\  
ENet + T1 (ours)                           & \textbf{0.652 $\pm$ 0.005}      & \textbf{0.959 $\pm$ 0.001} & 0.676 $\pm$ 0.007       &   0.965 $\pm$ 0.001 &                               \\  
ENet + T2 (ours)                              & 0.635 $\pm$ 0.003      & 0.958 $\pm$ 0.001     & 0.599 $\pm$ 0.009       &      \textbf{0.967 $\pm$ 0.001} &                            \\  
ENet + T3 (ours)                               & 0.634 $\pm$ 0.004      & 0.953 $\pm$ 0.001 & 0.648 $\pm$ 0.008       &       0.941 $\pm$ 0.001 &                           \\ \hline
MobileNetV2      & 0.540 $\pm$ 0.003       & 0.921 $\pm$ 0.002 & 0.516 $\pm$ 0.009       &     0.945 $\pm$ 0.001        & \multirow{4}{*}{2.2}                      \\  
MobileNetV2 + T1 (ours)                         & 0.595 $\pm$ 0.004      & 0.932 $\pm$ 0.002 & \textbf{0.684 $\pm$ 0.006}       &    0.952 $\pm$ 0.001 &                              \\  
MobileNetV2 + T2 (ours)                        & 0.590 $\pm$ 0.006       & 0.927 $\pm$ 0.002 & 0.678 $\pm$ 0.003       &     0.949 $\pm$ 0.001 &                             \\  
MobileNetV2 + T3 (ours)                         & 0.589 $\pm$ 0.002      & 0.924 $\pm$ 0.001 & 0.679 $\pm$ 0.005       &      n/a &                            \\ \hline
ResNet18         & 0.464 $\pm$ 0.008      & 0.934 $\pm$ 0.001     & 0.435 $\pm$ 0.005       &        0.933 $\pm$ 0.001 & \multirow{4}{*}{11.2}                         \\  
ResNet18 + T1 (ours)                           & 0.508 $\pm$ 0.004      & 0.943 $\pm$ 0.001 & 0.582 $\pm$ 0.008       &    0.939 $\pm$ 0.001 &                              \\  
ResNet18 + T2 (ours)                            & 0.491 $\pm$ 0.004      & 0.946 $\pm$ 0.001 & 0.551 $\pm$ 0.005       &     0.941 $\pm$ 0.001    &                         \\  
ResNet18 + T3 (ours)                            & 0.508 $\pm$ 0.006      & 0.935 $\pm$ 0.001 & 0.450 $\pm$ 0.009        &     0.934 $\pm$ 0.001 &                             \\ \hline
\end{tabular}
\label{table_core} 
\end{table*}

\begin{figure}[t]
\centering
\includegraphics[width=\columnwidth]{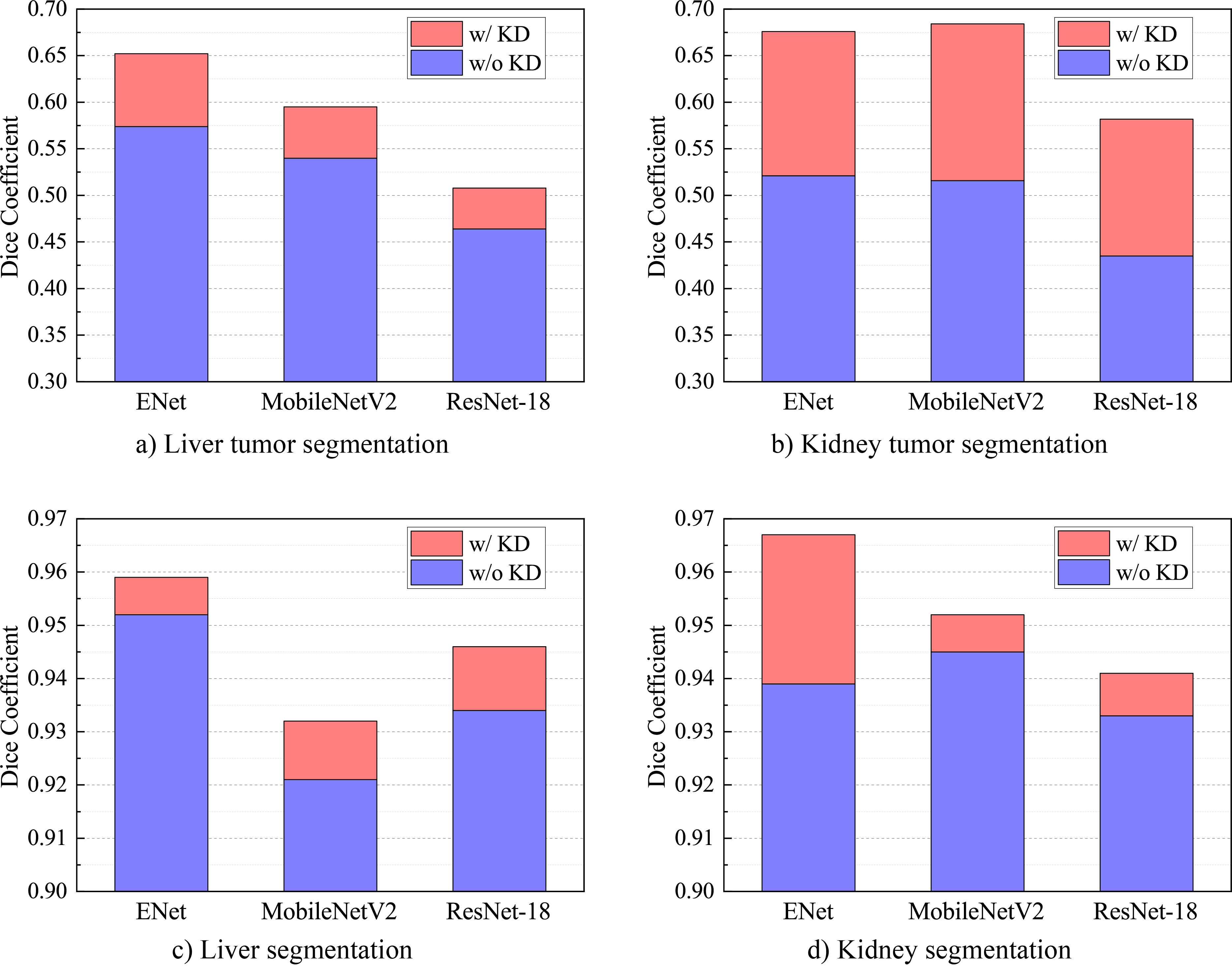}
\caption{Intuitive bar graphs of the effects of the knowledge distillation method we proposed. The promotion represented in pink is the maximum that we picked from our repeated experiments on both LiTS and KiTS19. Note that the start values of the vertical axis are different as the different difficulties of the corresponding tasks.}
\label{figure_promote}
\end{figure}

\begin{figure}[t]
\centering
\includegraphics[width=0.95\columnwidth]{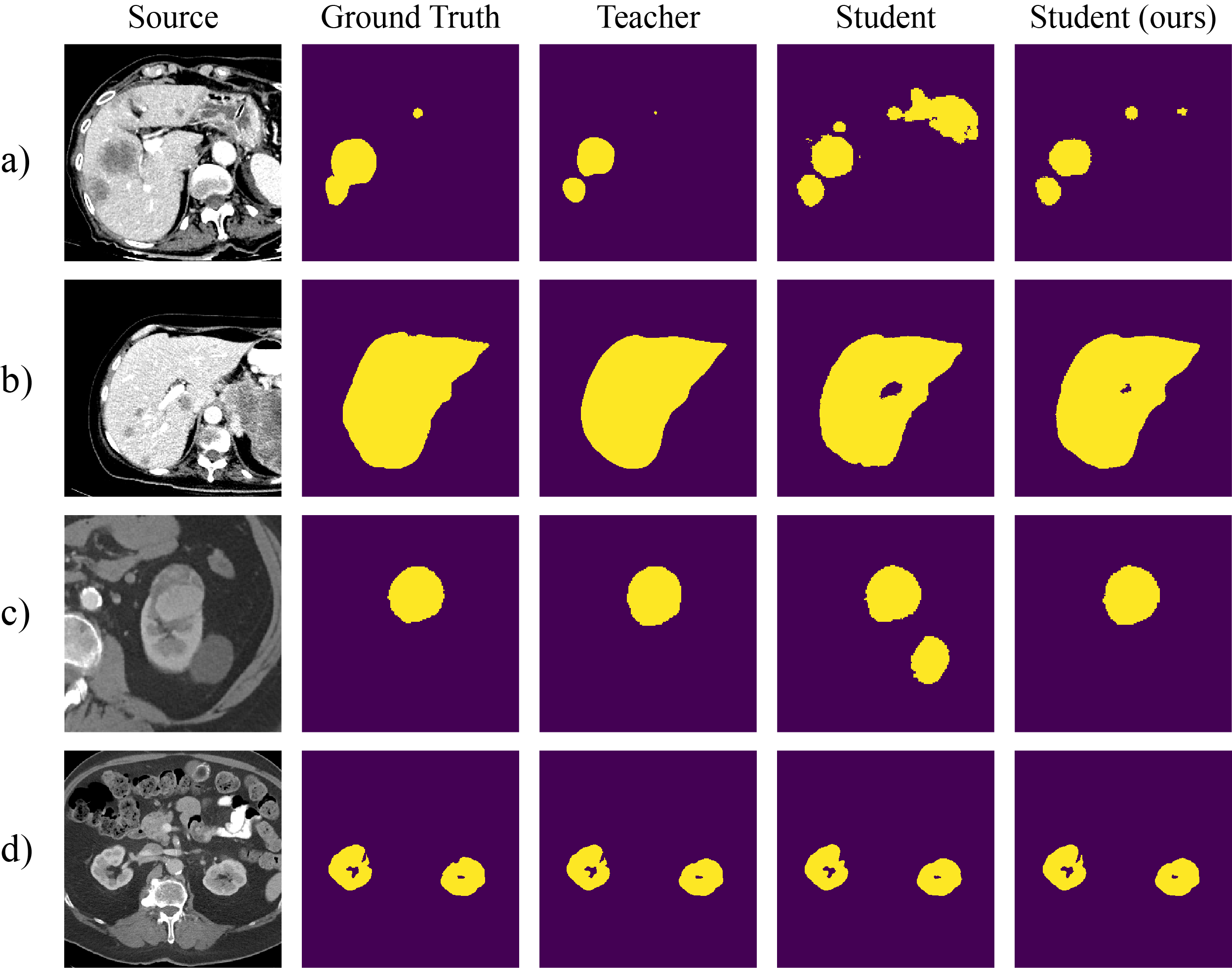}
\caption{Four representative segmentation results from our experiments: a) liver tumor; b) liver;c) kidney tumor; d) kidney. The teacher network is RA-UNet and the student network is ENet. As the pixel-level segmentation maps, we denote the background area as the purple region and the objective area as the yellow region.}
\label{figure_lits}
\end{figure}

\begin{figure*}[t]
\centering
\includegraphics[width=0.95\textwidth]{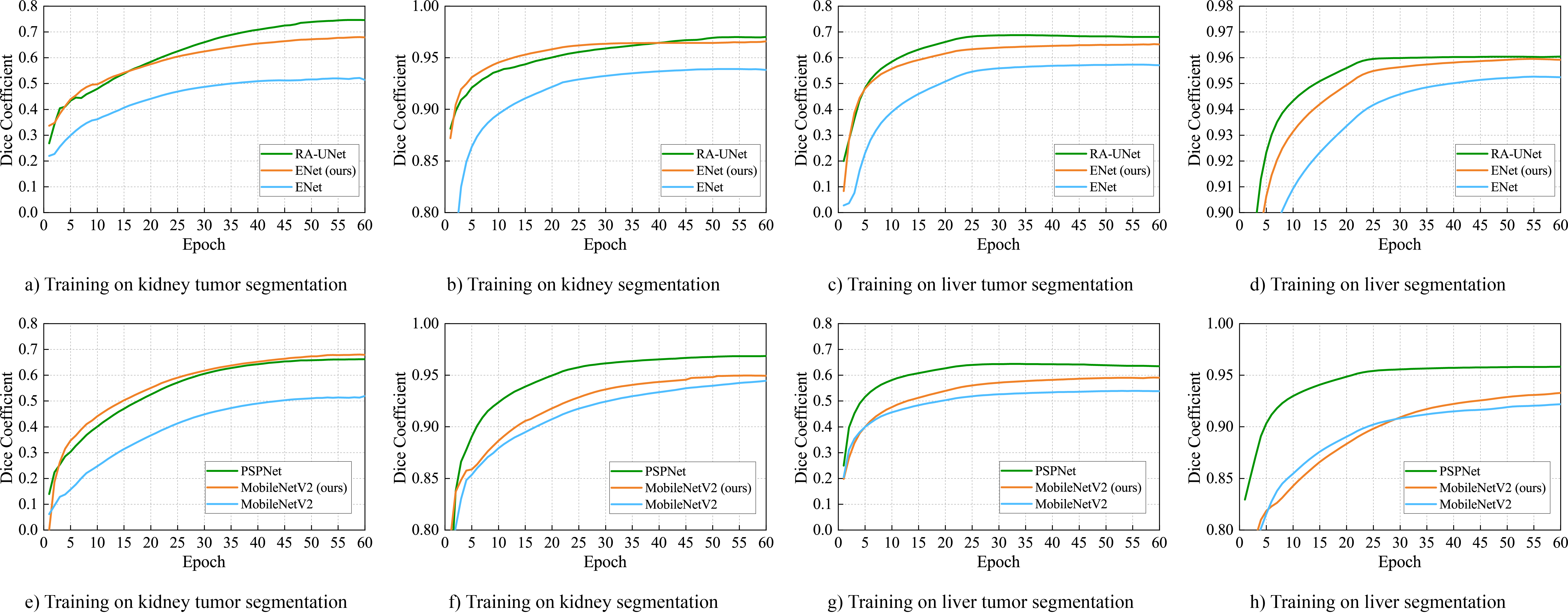}
\caption{Validation trend lines of the training process with our knowledge distillation methods. Note that they were painted in their own training processes as they were trained and updated separately. We coordinate them by the time measurement of epochs and evaluate their performance using the dice coefficient. There is a subtlety that we adjust the starting point of the horizontal axis of each chart to make it more intuitive.}
\label{figure_train}
\end{figure*}

\begin{table}[t]
\centering
\setlength{\tabcolsep}{5.7pt}
\renewcommand\arraystretch{1.4}
\caption{The rank of contemporary methods on liver and kidney tumor segmentation tasks. All networks are arranged in ascending order of the number of the parameters. The \underline{underlined method} is the teacher network of our ENet}
\begin{tabular}{c|c|c|c|c}
    \hline
    Method & \makecell[c]{\#Params \\ (M)} &  \makecell[c]{FLOPs\\(G)} & \makecell[c]{Liver Tumor \\ Dice} & \makecell[c]{Kidney Tumor \\ Dice} \\
    \hline
    ESPNet & 0.183 & 1.23 & 0.575 $\pm$ 0.006 & 0.462 $\pm$ 0.009 \\
    ENet & 0.353 & 2.03 & 0.574 $\pm$ 0.005 & 0.521 $\pm$ 0.015 \\
    MobileNetV2 & 2.2 & 19.14 & 0.540 $\pm$ 0.003 & 0.516 $\pm$ 0.009 \\
    ResNet-18 & 11.2 & 10.66 & 0.464 $\pm$ 0.008 & 0.435 $\pm$ 0.005 \\
    UNet++ & 20.6 & 620.04 & 0.669 $\pm$ 0.003 & 0.644 $\pm$ 0.007 \\
    \underline{RA-UNet} & 22.1 & 24.81 & 0.685 $\pm$ 0.004 & 0.745 $\pm$ 0.003 \\
    UNet & 34.5 & 293.83 & 0.658 $\pm$ 0.008 & 0.585 $\pm$ 0.010 \\
    PSPNet & 46.7 & 207.18 & 0.640 $\pm$ 0.005 & 0.659 $\pm$ 0.007 \\
    DeeplabV3+ & 56.8 & 272.48 & 0.641 $\pm$ 0.004 & 0.613 $\pm$ 0.012 \\
    ENet (ours) & 0.353 & 2.03 & 0.652 $\pm$ 0.005 & 0.676 $\pm$ 0.007 \\
    \hline
\end{tabular}
\label{table_rank}
\end{table}

\begin{figure}[t]
\centering
\includegraphics[width=0.95\columnwidth]{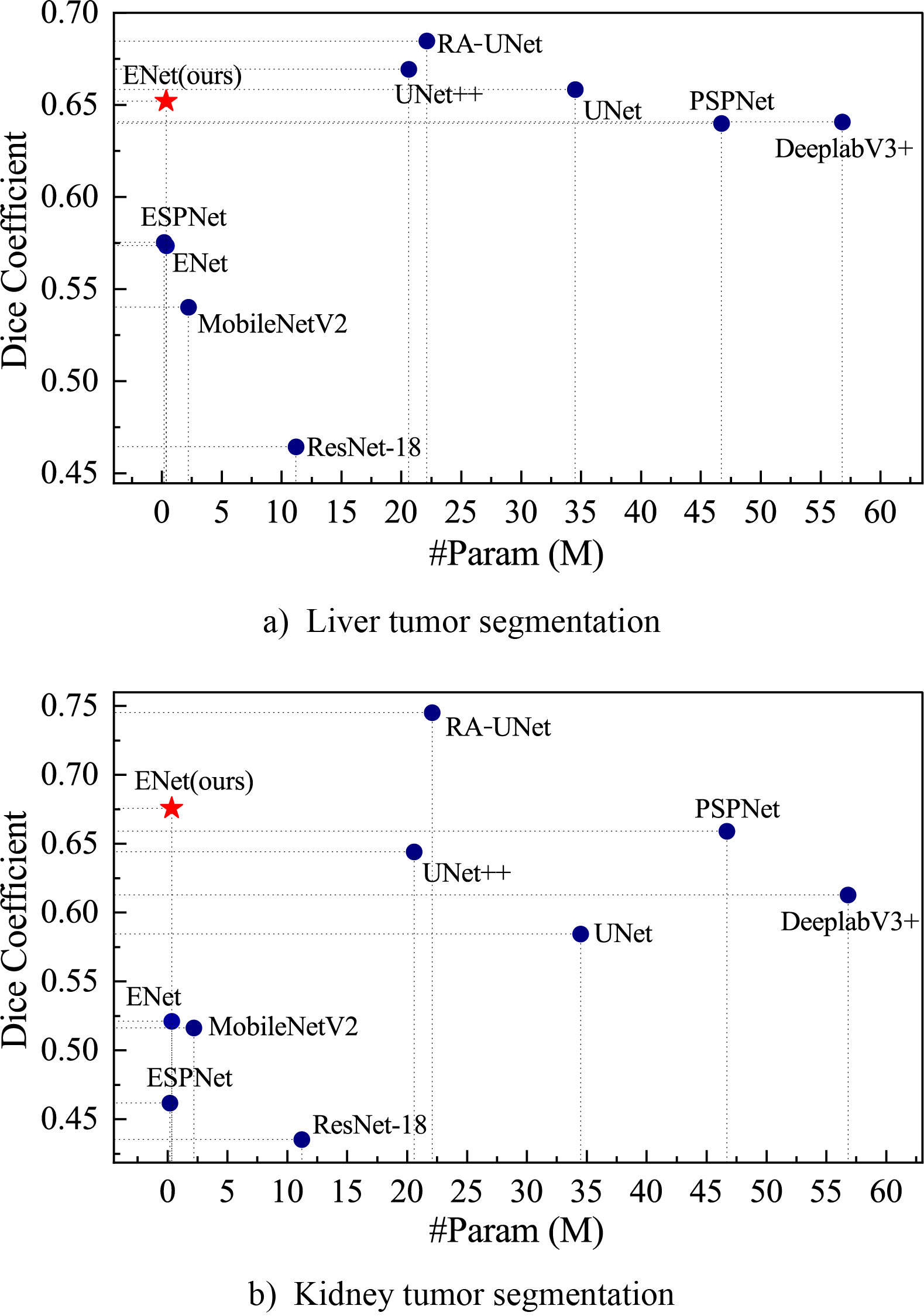}
\caption{The scatter diagrams of the segmentation capability of contemporary methods. The ideal method should be infinitely closer to the upper left corner.}
\label{figure_rank}
\end{figure}

\begin{table}[t]
\centering
\setlength{\tabcolsep}{10pt}
\renewcommand\arraystretch{1.4}
\caption{Comparison with other knowledge distillation methods on both LiTS and KiTS19. We fix the student and the teacher when using different distillation methods.}
\begin{tabular}{lcc}
    \hline
    Teacher       & \multicolumn{2}{c}{RA-UNet}                                        \\
    Student       & \multicolumn{2}{c}{ENet}                                           \\ \hline
                  & \multicolumn{1}{c}{Liver Tumor} & \multicolumn{1}{c}{Kidney Tumor} \\ \hline
    Teacher       & 0.685 $\pm$ 0.004                 & 0.745 $\pm$ 0.003                  \\
    Student       & 0.574 $\pm$ 0.005                 & 0.521 $\pm$ 0.015                  \\ \hline
    AT \cite{at}            & 0.640 $\pm$ 0.006                 & 0.650 $\pm$ 0.008                   \\
    PA \cite{skd}          & 0.618 $\pm$ 0.004                 & 0.535 $\pm$ 0.009                  \\
    SKD \cite{skd}          &  0.639 $\pm$ 0.009                 &  0.549 $\pm$ 0.009                  \\
    MIMIC \cite{mkd}        & 0.628 $\pm$ 0.001                 & 0.546 $\pm$ 0.009                  \\
    LOCAL \cite{local}       & 0.637 $\pm$ 0.003                 & 0.533 $\pm$ 0.010                   \\
    SPKD \cite{spkd}       & 0.635 $\pm$ 0.002                 & 0.602 $\pm$ 0.009                   \\
    IFVD \cite{ifvd}       & 0.640 $\pm$ 0.005                 & 0.580 $\pm$ 0.014                   \\
     \hline
    EMKD (ours) & \textbf{0.652 $\pm$ 0.005}                 & \textbf{0.676 $\pm$ 0.007}                  \\ \hline
    \end{tabular}
\label{table_kd}
\end{table}

\begin{table*}[t]
\centering
\setlength{\tabcolsep}{10pt}
\renewcommand\arraystretch{1.3}
\caption{The effectiveness of the components of our methods on the dataset LiTS. It should be noted that the score of Dice is the main measurement as the intuition of segmentation capability, while the scores of VOE and RVD are also given by us to enable readers to have a more comprehensive understanding of these components.}
\begin{tabular}{l|ccc|ccc}
    \hline
    \multicolumn{1}{c|}{\multirow{2}{*}{Method}} & \multicolumn{3}{c|}{Liver Tumor}                                              & \multicolumn{3}{c}{Liver}                                                    \\ \cline{2-7} 
    \multicolumn{1}{c|}{}                        & \multicolumn{1}{c}{Dice} & \multicolumn{1}{c}{VOE} & \multicolumn{1}{c|}{RVD} & \multicolumn{1}{c}{Dice} & \multicolumn{1}{c}{VOE} & \multicolumn{1}{c}{RVD} \\ \hline
    Teacher: RA-UNet                             & 0.685$\pm$0.004          & 0.204$\pm$0.013         & -0.083$\pm$0.027         & 0.960$\pm$0.001          & 0.051$\pm$0.002         & 0.024$\pm$0.003         \\ \hline
    Student: ENet                                & 0.574$\pm$0.005          & 0.238$\pm$0.018         & -0.064$\pm$0.046         & 0.956$\pm$0.001          & 0.057$\pm$0.002         & 0.027$\pm$0.003         \\
    + PMD                                        & 0.639$\pm$0.005          & 0.294$\pm$0.023         & 0.011$\pm$0.072          & 0.959$\pm$0.001          & 0.053$\pm$0.001         & 0.026$\pm$0.002         \\
    + IMD                                        & 0.645$\pm$0.003          & 0.273$\pm$0.017         & 0.024$\pm$0.052          & 0.958$\pm$0.001          & 0.054$\pm$0.001         & 0.025$\pm$0.003         \\
    + RAD                                        & 0.628$\pm$0.003          & 0.300$\pm$0.018         & 0.283$\pm$0.064          & 0.958$\pm$0.001          & 0.054$\pm$0.002         & 0.025$\pm$0.004         \\
    + PMD + IMD                                  & 0.646$\pm$0.004          & 0.318$\pm$0.027         & 0.185$\pm$0.069          & 0.959$\pm$0.001          & 0.055$\pm$0.003         & 0.024$\pm$0.006         \\
    + PMD + RAD                                  & 0.644$\pm$0.005          & 0.312$\pm$0.019         & 0.125$\pm$0.055          & 0.959$\pm$0.001          & 0.054$\pm$0.004         & 0.024$\pm$0.008         \\
    + IMD + RAD                                  & 0.642$\pm$0.003          & 0.256$\pm$0.015 & 0.005$\pm$0.056  & 0.959$\pm$0.001          & 0.053$\pm$0.001 & 0.024$\pm$0.002 \\
    + PMD + IMD + RAD                            & \textbf{0.652$\pm$0.005} & 0.231$\pm$0.036         & -0.092$\pm$0.074         & \textbf{0.959$\pm$0.001} & 0.071$\pm$0.003         & 0.024$\pm$0.009         \\ \hline
    \end{tabular}
\label{table_comp_li} 
\end{table*}

\begin{table*}[t]
\centering
\setlength{\tabcolsep}{10pt}
\renewcommand\arraystretch{1.3}
\caption{The effectiveness of the components of our methods on the dataset KiTS19. The setup is the same as Table \ref{table_comp_li}}
\begin{tabular}{l|ccc|ccc}
\hline
\multicolumn{1}{c|}{\multirow{2}{*}{Method}} & \multicolumn{3}{c|}{Kidney Tumor}                                             & \multicolumn{3}{c}{Kidney}                                                   \\ \cline{2-7} 
\multicolumn{1}{c|}{}                        & \multicolumn{1}{c}{Dice} & \multicolumn{1}{c}{VOE} & \multicolumn{1}{c|}{RVD} & \multicolumn{1}{c}{Dice} & \multicolumn{1}{c}{VOE} & \multicolumn{1}{c}{RVD} \\ \hline
Teacher: RA-UNet                             & 0.745$\pm$0.003          & 0.205$\pm$0.008         & 0.007$\pm$0.020          & 0.970$\pm$0.001           & 0.026$\pm$0.001         & -0.006$\pm$0.002        \\ \hline
Student: ENet                                & 0.521$\pm$0.015          & 0.248$\pm$0.036         & -0.189$\pm$0.080         & 0.939$\pm$0.001          & 0.039$\pm$0.003         & -0.022$\pm$0.005        \\
+ PMD                                        & 0.608$\pm$0.007          & 0.248$\pm$0.025         & -0.082$\pm$0.052         & 0.946$\pm$0.001          & 0.032$\pm$0.002         & -0.019$\pm$0.004        \\
+ IMD                                        & 0.653$\pm$0.006          & 0.204$\pm$0.013         & -0.083$\pm$0.037         & 0.950$\pm$0.002          & 0.031$\pm$0.001         & -0.020$\pm$0.003        \\
+ RAD                                        & 0.646$\pm$0.008          & 0.232$\pm$0.019         & -0.005$\pm$0.050         & 0.948$\pm$0.001          & 0.030$\pm$0.001         & -0.022$\pm$0.003        \\
+ PMD + IMD                                  & 0.669$\pm$0.007          & 0.212$\pm$0.020         & -0.052$\pm$0.047         & 0.959$\pm$0.001          & 0.031$\pm$0.001         & -0.013$\pm$0.003        \\
+ PMD + RAD                                  & 0.667$\pm$0.005          & 0.199$\pm$0.013         & -0.065$\pm$0.032         & 0.954$\pm$0.002          & 0.033$\pm$0.002         & -0.014$\pm$0.005        \\
+ IMD + RAD                                  & 0.670$\pm$0.004          & 0.193$\pm$0.015 & -0.023$\pm$0.042  & 0.961$\pm$0.001          & 0.032$\pm$0.001 & -0.011$\pm$0.002 \\
+ PMD + IMD + RAD                            & \textbf{0.676$\pm$0.007} & 0.184$\pm$0.008         & -0.040$\pm$0.021         & \textbf{0.965$\pm$0.001} & 0.029$\pm$0.002         & -0.008$\pm$0.004        \\ \hline
\end{tabular}
\label{table_comp_ki} 
\end{table*}

\begin{table}[t]
\centering
\setlength{\tabcolsep}{7pt}
\renewcommand\arraystretch{1.3}
\caption{The experimental results of influences of the component weights represented by hyper-parameters $\alpha$, $\beta_1$,and $\beta_2$ in Eq. \ref{eq_all}. As shown in the first two rows, the training process will be equivalent to training the original regular segmentation network when these weights are set to 0.}
\begin{tabular}{c|ccc|c}
\hline
\multirow{2}{*}{Method}        & \multicolumn{3}{c|}{Weight of Components} & \multirow{2}{*}{Kidney Tumor Dice} \\ \cline{2-4}
                                & $\alpha$   & $\beta_1$  & $\beta_2$  &                                    \\ \hline
Teacher: RA-UNet               & 0          & 0          & 0          & 0.745 $\pm$ 0.003                  \\
Student: ENet                  & 0          & 0          & 0          & 0.521 $\pm$ 0.015                  \\ \hline
\multirow{5}{*}{\makecell[c]{ENet \\+\\ EMKD (ours)}} & 0.1        & 0.9        & 0.9        & \textbf{0.676 $\pm$ 0.007}                  \\
                                & 0.2        & 0.9        & 0.9        & 0.672 $\pm$ 0.015                  \\
                                & 0.1        & 1.8        & 0.9        & 0.675 $\pm$ 0.006                  \\
                                & 0.1        & 0.9        & 1.8        & 0.675 $\pm$ 0.009                  \\
                                & 0.1        & 1.8        & 1.8        & 0.673 $\pm$ 0.011                  \\
                                \hline
\end{tabular}
\label{table_param}
\end{table}

\begin{figure}[t]
\centering
\includegraphics[width=\columnwidth]{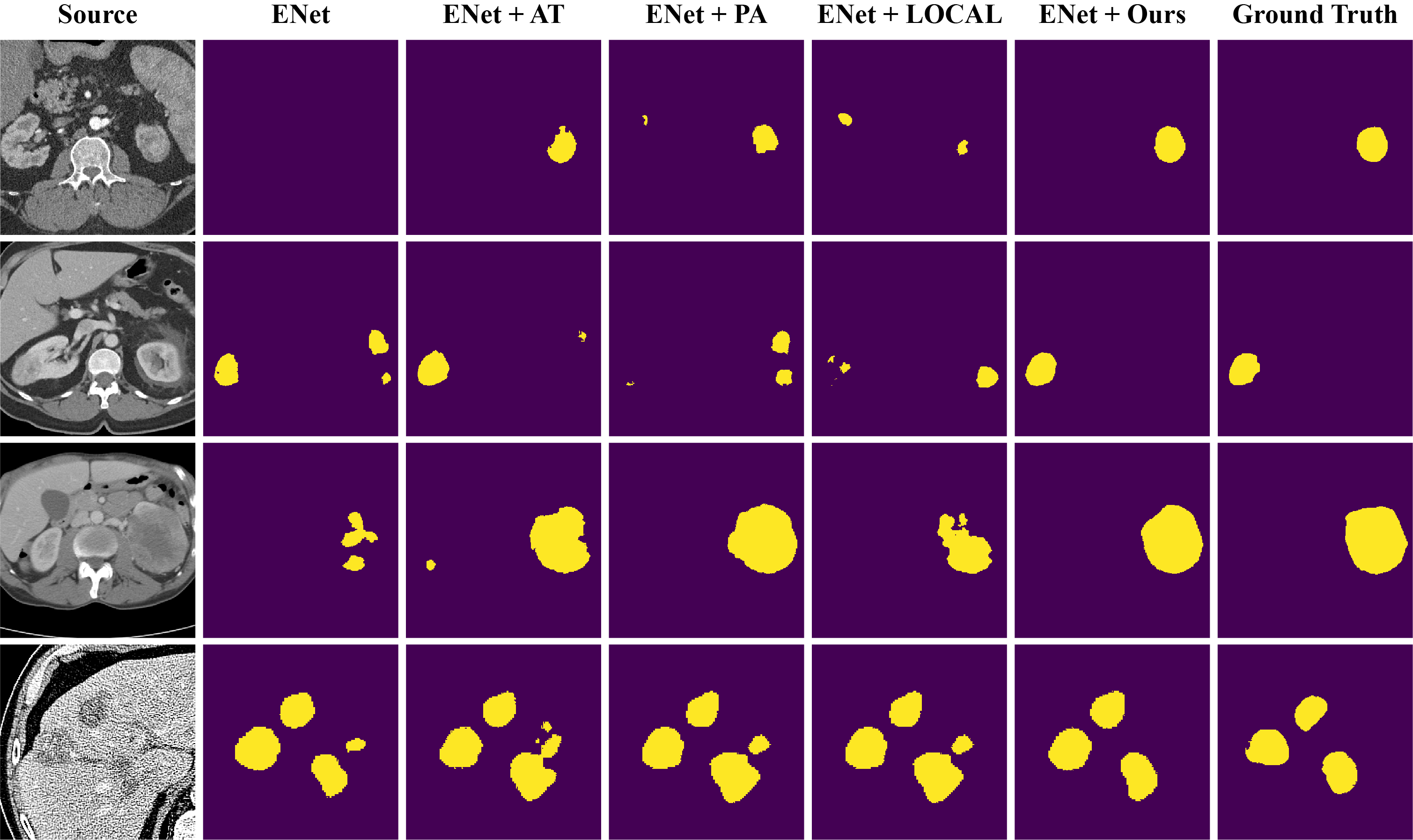}
\caption{Visualization of the prediction results of knowledge distillation methods. We adopt the constant teacher network RA-UNet for all distillation methods here. As the pixel-level segmentation maps, we denote the background area as the purple region and the objective area as the yellow region.}
\label{figure_vis}
\end{figure}

In this paper, we conduct the ablation study through experiments of various perspectives. First, To demonstrate the power of the distillation method proposed by us, we train and verify our architecture by distilling from the different teacher and student networks. Several state-of-the-art segmentation networks where some of them are tailored for medical imaging are adopted as the teacher networks, such as RA-UNet \cite{rau}, PSPNet \cite{psp}, and UNet++ \cite{unetp}. We also select some commonly applied lightweight networks such as ENet \cite{enet},  MobileNetV2 \cite{mv2}, and ResNet-18 \cite{res} as the student networks. Then, we list piles of contemporary epidemic networks regardless of their body type to show the advantages and the position of our approach in modern methods. We also demonstrate that our method can reach state-of-the-art performance in distilling through the experiments of comparing with other knowledge distillation methods. In the end, we take further ablation consideration about the three distillation modules in our architecture and the hyper-parameters in Eq. \ref{eq_all}.

\subsubsection{Primary Results}

As the core part of this ablation study, we apply our distillation architecture on multiple pairs of teacher and student networks and verify on both LiTS and KiTS19. There is an obstacle when distilling the intermediate features that the changes in the size of features in the process are different as the inconsistent number of up and downsampling layers. To solve this, we uniformly extract the first and the last embedding pairs of the same size which can be found as possible as the representative low-level and high-level feature pairs, then feed them to our distillation modules. 

We adopt commonly applied medical image segmentation models RA-UNet, PSPNet, and UNet++ as our teachers in this part of experiments. Table \ref{table_core} presents the results. What can be observed is that all student networks are able to reach higher performance by learning from any teacher network which is stronger than them through our knowledge distillation method. It is also willing to see that our method is effective for all the segmentation tasks. The student network ENet, MobileNetV2, and ResNet-18 embrace the maximal improvement of 13.6\% (0.078), 10\% (0.055), and 9.5\% (0.044) in dice coefficient score for the liver tumor segmentation respectively. The three students also gain the promotion up to the percentage of 0.7\% (0.007), 1.1\% (0.011) and 1.2\% (0.012) for the liver segmentation. Our method has even more amazing effects on the improvement of kidney tumor segmentation. The most visible promotion value \textbf{0.168} of dice score is contributed by the teacher RA-UNet and the student MobileNetV2. In other words, the performance of MobileNetV2 on kidney tumor segmentation can be elevated in a percentage of \textbf{32.6\%}. The most excellent student is ENet for kidney segmentation. It reaches the score of 0.967 after finishing the learning from the teacher PSPNet. Figure \ref{figure_promote} presents the power of our method in an intuitive way. 

Obviously, some students can reach the performance which is very close to the level of the teachers in all the four segmentation tasks. Figure \ref{figure_lits} presents some visualized cases from LiTS and KiTS19. It can be observed that our method can not only correct the mistakes made by students but also drive their segmentation results close to the ground truth. 

More than that, the method we proposed can also accelerate the speed of convergence in most cases. Figure \ref{figure_train} illustrates the training process of some experiments. We applied a validation strategy of recording dice coefficient scores after the end of every training epoch. As these trend lines presented in the figure, our method is skilled in improving the students who should have performed poorly in training to almost the same level as their teacher.

\subsubsection{Contemporary Rank}

The mission of knowledge distillation is to make the networks lighter or improve the performance of lightweight networks. To show our level in contemporary academia clearly, the lightweight network ENet distilled from RA-UNet using our method is ranked among nominated models as Table \ref{table_rank} and Figure \ref{figure_rank}. The candidates contain not only the networks mentioned above but also some epidemic segmentation methods such as ESPNet \cite{esp}, UNet \cite{unet}, and DeeplabV3+ \cite{deep}. The FLOPs calculated by feeding in a constant input of the size $384\times384$ are also listed in Table \ref{table_rank} to interpret the computational complexity of the models.

As presented, the student network ENet distilled by our method achieves the 4th in liver tumor dice and surpasses some state-of-the-art segmentation models such as PSPNet and DeeplabV3+. The more exciting thing is that our student network reaches the dice coefficient score of 0.676 and beats all other models except RA-UNet on the kidney tumor segmentation task as illustrated in Figure \ref{figure_rank}-b. Do not forget to check the size of these models, we always retain the very few parameters of the original student model. There is no doubt that it is hard to find an off-the-shelf lightweight network that possesses the capability to compare with the network distilled by our method.

\subsubsection{Comparison with Other Knowledge Distillation Methods}
It is necessary to compare our method with other knowledge distillation approaches. We embrace some recent methods such as PA \cite{skd}, MIMIC \cite{mdk}, LOCAL \cite{local}, and IFVD \cite{ifvd}, although the corresponding research on segmentation problems is still scarce. We also implement two commonly applied methods, AT \cite{at} and SPKD \cite{spkd}, while they are not for segmentation problems when proposing. We conduct this part of experiments with constant teacher RA-UNet and student network ENet and extract the features in the process in the same position of the networks to guarantee that the different distillation methods are carried out in the same environment.

Table \ref{table_kd} shows the results of the comparison. Obviously, our method dubbed EMKD takes the crown of the competition of knowledge distillation in both two tasks and holds remarkable advantages in kidney tumor segmentation. We further visualize their performance on the same inputs as represented in Figure \ref{figure_vis}.

\subsubsection{The Effectiveness of Distillation Components}
As the last part of our experiments, we verify the effectiveness of all the components, including the modules of Prediction Maps Distillation (PMD), Importance Maps Distillation (IMD), and Region Affinity Distillation (RAD) in our architecture. Table \ref{table_comp_li} and Table \ref{table_comp_ki} show the results on LiTS and KiTS19 with the dice coefficient score and another two evaluative metrics, VOE and RVD. One can drive from it immediately that every component has a positive effect on the performance of the student network. The last row of each table further demonstrates that our architecture assembled with the three modules can reach the best performance. Obviously, our novel modules IMD and RAD play key roles in the final distillation method. Take the results on kidney tumor segmentation in Table \ref{table_comp_ki} as an example. The base distillation module PMD gives a promotion of 0.087 of dice score, from 0.521 to 0.608. The IMD and RAD modules further increase 0.068 of dice score, from 0.608 to 0.676. It needs to be emphasized that the room for distillation is limited by the gap of performance between the teacher and student network. Theoretically, it is hard to get a remarkable improvement for existing knowledge distillation methods when the gap is tiny, such as the experimental results of liver segmentation in Table \ref{table_comp_li}.

We also demonstrate the insensitivity of our method to the hyper-parameters. Given the weights $\alpha$, $\beta_1$, and $\beta_2$ for the three modules in the total loss function Eq. \ref{eq_all}, we initialize them by the experimented optimum values of 0.1, 0.9, and 0.9. As represented in Table \ref{table_param}, only some relatively slight performance drop could be perceived after doubling these values, and the influences of adjusting $\beta_1$ or $\beta_2$ solely can be ignored. Thus, we prefer to alter the values of $\beta_1$ and $\beta_2$ simultaneously in practice.

\section{Discussion}
This work is supposed to be the pioneer that systematically constructs a knowledge distillation architecture for medical image segmentation problems. The implanted three distillation modules in our architecture orderly take charge of guaranteeing the basic effectiveness of the knowledge transfer, paying attention to the important neurons, and excavating the inter-class semantic information. To the best of our knowledge, the proposed novel module RAD is the first distillation method tailored for medical image segmentation. Different from prior works \cite{xray}-\cite{mono} on medical image, this method is supposed to be a pioneering example to consider utilizing the relationships among the different semantic classes in a contrastive way. The clever twist is that this method effectively steers clear of the ambiguous boundary problems when facing medical image segmentation tasks. The experimental results in this paper demonstrate the flexibility of our method. Theoretically, our architecture allows any convolutional networks that conform with the encoder-decoder structure to be the student and teacher networks. In addition, the compatibility of heterogeneous network architecture between teacher and student models is also guaranteed.

The distillation methods in our architecture are designed to be conveniently reproduced and escalated. All roads lead to Rome. For instance, any one of the three distillation modules would be upgraded by future researches. The modules can also be replaced with other distillation methods when needed. Such as adopting IFVD \cite{ifvd} rather than IMD to cooperate with RAD to encode the inter-class and inner-class semantic information at the same time. Moreover, the number of distillation modules can be unlimited. In other words, it is worth trying to append one or more new knowledge distillation methods after RAD to reach better performance by squeezing the rest distillation room. In practice, our method can also be applied in other semantic segmentation problems which require the distillation mechanism. Since the structural knowledge distillation \cite{skd} is successfully verified on the well-known and challenging semantic segmentation datasets Cityscapes \cite{city} and ADE20K \cite{ade}, the core methodology in our RAD module that tries to transfer the relationship information between different classified regions also has the potential to be a novel and effective distillation way to tackle general segmentation problems.

Some interesting experimental results can be observed in Table \ref{table_core}. In the task of kidney tumor segmentation, the performances of MobileNetV2 reach the dice score of 0.678 and 0.679 after finishing distillation from the teacher network PSPNet and UNet++, which surpasses the performance of the two teachers with the dice score of 0.659 and 0.644. We suppose that this phenomenon implies that our architecture can guide the student network to understand semantic information better. With the evidence in the table that our method performs the best in the kidney tumor segmentation task, the underlying reason may be that the inter-class semantic information is more richly excavated in this data distribution. Of course, the above suppositions need to be verified in future works.

Our work can be further improved in the future. When discussing medical image processing, it is reasonable to consider the applicability in 3D scenarios. However, there are still several major issues to be resolved. First, most existing knowledge distillation methods, including our work, are designed to utilize intermediate feature maps efficiently. For 3D segmentation tasks, the computational complexity and storage usage tend to be impractical as the distillation methods often require frequent calculations on the 3D feature maps of both teacher and student networks. Second, to transfer meaningful and effective information is more challenging as the ratio of the area of the objective region to the background region is commonly smaller in 3D scenarios. Moreover, not all medical image datasets are suitable to apply 3D networks. Take our experiments on LiTS and KiTS19 as the example. The $z$ dimension will be disappeared in the convolution process in that the minimum number of the slices is 42 and 20, respectively. Although our architecture can be readily extended and implemented in 3D scenarios, a well-planned scheme that systematically considers the above issues still requires many new ideas and workloads, enough to be published as another single paper.

Another way of improvement is to accommodate multiple models in our knowledge distillation structure. Some related researches have aroused recently, such as distillation from multi-teacher to single-student \cite{mt}, and from single-teacher to multi-student \cite{ms}. However, it is still challenging to integrate the feature maps of different sizes from more than one teacher or student network and then feed them to the embedded distillation functions, as our architecture consists of three I/O standardized knowledge distillation modules. Therefore, we will devote ourselves to cope with the above works in the future.

\section{Conclusion}
In this paper, we have proposed a novel distillation architecture tailored for the medical image segmentation problem. We have also demonstrated that our method has the ability to transfer structural information from cumbersome networks to lightweight networks through a series of convictive experiments. After distilling, the lightweight network got a remarkable improvement and reached a performance comparable to the state-of-the-art cumbersome networks. We believe this work will help to pave the way for further researches, especially those focusing on both the medical image segmentation problem and the knowledge distillation technology. We hope that this paper can ignite a mass fervor for researchers that pay close attention to the field.

\bibliographystyle{IEEEtran.bst}

\end{document}